\documentclass[12 pt, amssymb,prb,showpacs]{revtex4}
\usepackage{epsfig}
\usepackage{dcolumn}
\usepackage{amsmath}
\usepackage[a4paper,dvips]{geometry}
\begin{document}
\title{\bf Anharmonic behavior in Microwave-driven resistivity oscillations in Hall bars}
\author{Jes\'us I\~narrea }
 \affiliation {$^1$Escuela Polit\'ecnica
Superior,Universidad Carlos III,Leganes,Madrid,Spain  and \\
$^2$Instituto de Ciencia de Materiales, CSIC,
Cantoblanco,Madrid,28049,Spain.}
\date{\today}
%%%%%%%%%%%%%%%%%%%%%%%%%%%%%%%%%%%%%%%%%%%%%%%%%%%%%%%%%%%%%%%%%%
%%%%%%%%%%%%
\begin{abstract}
We analyzed the magnetoresistivity of a two-dimensional electron
system excited by microwave radiation in a regime of high
intensities and low frequencies. In such a regime, recent
experiments show that different features appear  in the
magnetoresistivity response which suggest an anharmonic behavior.
These features consist mainly in distorted oscillations and new
resonance peaks at the subharmonics of the cyclotron frequency. We
follow the model of microwave-driven electron orbits motion which
become anharmonic when the ratio of microwave intensity to microwave
frequency is large enough.
\end{abstract}
%%%%%%%%%%%%%%%%%%%%%%%%%%%%%%%%%%%%%%%%%%%%%%%%%%%%%%%%%%%%%%%%%%
%%%%%%%%%%%%
\maketitle
\newpage

Microwave-Induced Resistivity Oscillations
(MIRO)\cite{zudov,studenikin}  and Zero Resistance States
(ZRS)\cite{mani,zudov2} have been continuously attracting much
attention in the recent years. From the experimental standpoint,
different features and improvements are being implemented in the
corresponding experimental
configurations\cite{willett,mani2,zudov3,yang,mani3,smet}. Thus
experimentalist have been introducing changes in the orientations,
frequencies, polarizations and intensities of the different fields
involved. Accordingly the results obtained with these configurations
are getting more and more striking and difficult to explain. In some
cases totally unexpected outcomes have been obtained \cite{smet}.
All these experimental evidences and results establish real
challenges which can be regarded as crucial tests for the
theoretical models currently
available\cite{girvin,lei,ryzhii,rivera,andreev,ina,mirlin}.
Recently more experimental outcomes have joined the group of
striking experimental contributions\cite{doro,zudov4}. In this case
MIRO's are specifically studied in a regime of high MW intensities
and low MW frequencies ($w$). The remarkable results show distorted
profiles in the diagonal magnetoresistivity ($\rho_{xx}$)
oscillations and new resonance peaks at the subharmonics of the
cyclotron frequencies. These effects become more important when the
ratio MW power to MW frequency is larger and exceed a certain
threshold. It is also noticeable that the main oscillations are
shifted to lower magnetic field as the latter ratio increases.

In this letter we report on a theoretical explanation to this
unexpected behavior of $\rho_{xx}$ response obtained in regime of
large MW power to  $w$ ratio. We follow the model of MW driven
Larmor orbits\cite{ina} under this extreme regime. According to this
model, the electronic orbit guiding center performs under MW
excitation, an oscillating and harmonic motion with the same
frequency as MW. Thus, any change in the experimental set-up
affecting the electronic orbit motion, will be reflected in the
corresponding $\rho_{xx}$ response. A very direct example is the
experimental proposal regarding $\rho_{xx}$ response to bichromatic
MW radiation\cite{zudov3} and the corresponding theoretical
explanation\cite{ina4}. Following this model, if we increase the MW
power and decrease $w$, the amplitude of the Larmor orbit
oscillation becomes larger and larger making eventually the
oscillating motion $anharmonic$. Then the electrons in their Larmor
orbits become {\it driven anharmonic oscillators}. In this letter we
consider only the case of slightly driven anharmonic behavior. It is
well-known that a small amplitude driven anharmonic oscillating
system has the following characteristics: (1) the forced
oscillations contains harmonics that are not present in the driving
force; (2) subsidiary resonances occur at driving frequencies which
are subharmonic of the main resonance frequency; and  (3) the main
oscillations are moved to lower frequencies\cite{main}. All this
features are clearly present in the experimental results\cite{doro}.
A possible microscopic mechanism for the anharmonic oscillating
motion involves an interplay between a large oscillation amplitude
and the interaction between electrons in their MW-driven oscillating
Larmor orbits and the lattice ions.

%These interactions yield acoustic phonons and lattice deformations.
%For small amplitudes we expect small deformations and approximately
%symmetric interactions between electrons and lattice in the back and
%forth oscillating movement. This is characteristic of a harmonic
%behavior. However as the amplitude increases the lattice deformation
%gets larger giving rise to non-symmetric interactions in each part
%of the oscillation. The result is that the potential becomes
%anharmonic and the system shows a non-linear behavior. If the
%amplitude is large enough the system can be led to a chaotic regime.

 Following the MW driven Larmor orbits model, we
first obtain the exact expression of the electronic wave vector for
a two-dimensional electron system (2DES) in a perpendicular and
moderate magnetic field $B$,  and MW radiation\cite{ina,ina5}:
\begin{eqnarray}
&&\Psi(x,y,t)=\phi_{N}\left[(x-X-x_{cl}(t)),(y-y_{cl}(t)),t\right]\nonumber  \\
&&\times  exp \frac{i}{\hbar} \left[m^{*}\left(\frac{d
x_{cl}}{dt}x+\frac{d y_{cl}}{dt}y\right)+
\frac{m^{*}w_{c}(x_{cl}x-y_{cl}y)}{2}-\int_{0}^{t} {\it L} dt'\right]\nonumber  \\
&&\times\sum_{p=-\infty}^{\infty} J_{p}(A_{N}) e^{ipwt}
\end{eqnarray}
where  $\phi_{N}$ are analytical solutions for the Schr\"{o}dinger
equation with a two-dimensional (2D) parabolic confinement, known as
Fock-Darwin states. $X$ is the center of the orbit for the electron
spiral motion. $x_{cl}(t)$ and $y_{cl}(t)$ are the classical
solutions for a driven 2D harmonic oscillator. For MW radiation
polarized along the current direction (x-direction) and a harmonic
oscillation regime, the expression for $x_{cl}$ is given
by\cite{ina,ina5}:
\begin{equation}
x_{cl}(t)=\frac{e
E_{o}}{m^{*}\sqrt{(w_{c}^{2}-w^{2})^{2}+\gamma^{4}}}\cos wt
\end{equation}
where $\gamma$ is a sample dependent damping parameter which affects
dramatically the MW-driven electronic orbits movement. Along with
this movement there occur interactions between electrons and lattice
ions yielding acoustic phonons and producing a damping effect in the
electronic motion\cite{ina}. $E_{o}$ is the amplitude of the MW
field. ${L}$ is the classical lagrangian, and $J_{p}$ are Bessel
functions\cite{ina,ina5,ina6}.

In the present regime, $x_{cl}$ and $y_{cl}$ are the solutions for
the dynamics of a {\it 2D driven classical anharmonic oscillator}.
Since we do not know the exact nature of the anharmonic term in the
corresponding potential is impossible to solve analytically the
classical equation of motion\cite{balcou}. Nevertheless we can take
an alternative approach if we consider that the oscillating orbits
are only slightly anharmonic. Then, although not harmonic, the
system can be consider still periodic. As for any periodic function,
we can try to express $x_{cl}$ through a Fourier series and propose
a solution like:
\begin{equation}
x_{cl}(t)=\frac{A_{0}}{2}+\sum_{n=1}^{\infty}[A_{n}\cos(nw
t)+B_{n}\sin (nwt)]
\end{equation}
where $A_{0}$, $A_{n}$ and $B_{n}$ are the corresponding Fourier
coefficients.
\begin{figure}
\centering\epsfxsize=3.5in \epsfysize=4.0in
\epsffile{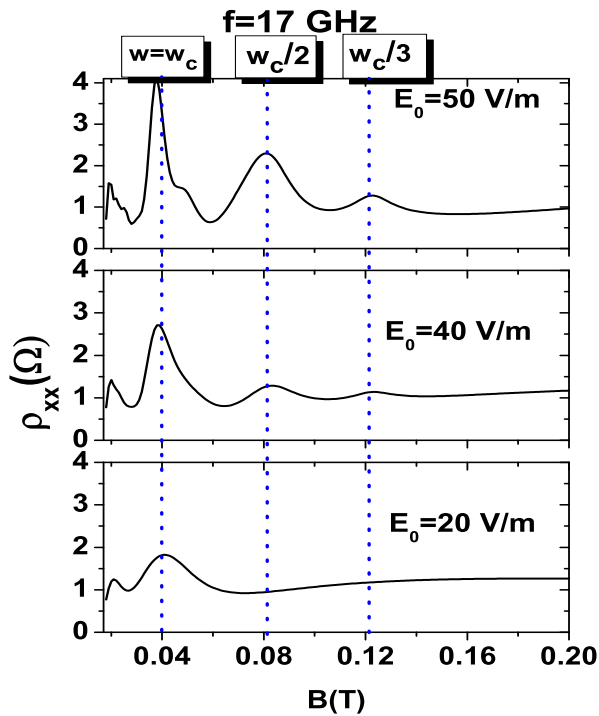} \caption{Calculated magnetoresistivity
$\rho_{xx}$ as a function of $B$ for different MW intensities. MW
frequency is $17 GHz$. Vertical dashed lines correspond to the
position of the cyclotron resonance ($w=w_{c}$) and its
subharmnonics. It can be observed very clearly the increasing
anharmonicity of $\rho_{xx}$ as the MW power is getting larger.
T=1K.}
\end{figure}
%\begin{figure} \centering\epsfxsize=3.5in
%\epsfysize=3.5in\epsffile{anarmonico2.eps} \caption{Same as in Fig.
%1: calculated $\rho_{xx}$ versus $B$ for increasing MW power and
%$w=17 GHz$. In this case all results are displayed in the same
%panel.}
%\end{figure}

 Now we introduce the scattering
suffered by the electrons due to charged impurities randomly
distributed in the sample. To proceed,  following the model
described in reference [16], firstly we calculate the
electron-charged impurity scattering rate $1/\tau$ (being $\tau$ the
scattering time). Secondly we find the average effective distance
advanced by the electron in every scattering jump, that in the case
of anharmonicity is given by:
\begin{equation}
\Delta X^{MW}=\Delta
X^{0}+A_{0}/2+\sum_{n}[A_{n}\cos(nw\tau)+B_{n}\sin (nw\tau)]
\end{equation}
where $\Delta X^{0}$ is the effective distance advanced when there
is no MW field present\cite{ina}. The longitudinal conductivity
$\sigma_{xx}$ can be calculated: $\sigma_{xx}\propto \int dE
\frac{\Delta X^{MW}}{\tau}(f_{i}-f_{f})$,  being $f_{i}$ and $f_{f}$
the corresponding distribution functions for the initial and final
states respectively and $E$ energy. To obtain $\rho_{xx}$ we use the
relation
$\rho_{xx}=\frac{\sigma_{xx}}{\sigma_{xx}^{2}+\sigma_{xy}^{2}}
\simeq\frac{\sigma_{xx}}{\sigma_{xy}^{2}}$, where
$\sigma_{xy}\simeq\frac{n_{i}e}{B}$ and $\sigma_{xx}\ll\sigma_{xy}$.
Finally we can express $\rho_{xx}$ as being proportional to a sum of
Fourier terms which reads:
\begin{equation}
\rho_{xx}\propto \sum_{n}[A_{n}\cos(nw\tau)+B_{n}\sin (nw\tau)]
\end{equation}
In order to obtain the Fourier terms in $\rho_{xx}$, we have carried
out a Fourier synthesis process. This process consists in
constructing the $\rho_{xx}$ form by adding together a fundamental
frequency (which corresponds to the harmonic case) and overtones of
different amplitudes. Since at this stage it is impossible to obtain
analytical expressions for the Fourier coefficients, we have
introduced phenomenologically the following ones:\\
 $A_{n}=\alpha_{n}\frac{e
E_{o}}{m^{*}\sqrt{(w_{c}^{2}-(nw)^{2})^{2}+\gamma^{4}}}=\alpha_{n}C_{n}$
and \\
$B_{n}=\beta_{n}\frac{e
E_{o}}{m^{*}\sqrt{(w_{c}^{2}-w^{2})^{2}+\gamma^{4}}}=\beta_{n}C_{1}$
where $\alpha_{n}$ and $\beta_{n}$ are anharmonicity terms. Their
values are getting larger as the anharmonicity increases.

In Fig.1,  we show the calculated $\rho_{xx}$,  as a function of $B$
for different MW intensities and $w=17 GHz$. Vertical dashed lines
are positioned at the cyclotron resonance ($w=w_{c}$) and its
subharmnonics.
%In Fig. 2, we present the same as in Fig.1, i.e.,
%calculated $\rho_{xx}$ vs $B$ for increasing MW power and $w=17
%GHz$. In this figure all results have been displayed in the same
%panel. Again vertical dashed lines indicate the position of the
%different resonance conditions.
For the three curves presented different developed expressions of
$\rho_{xx}$ have been used. Each one shows the increasing
anharmonicity for increasing $E_{0}/w$ ratio. As we said above, they
have been obtained through a Fourier synthesis process where the
anharmonicity coefficients have been introduced phenomenologically,
keeping the number of Fourier terms as small as possible. Thus for
the top panel of Fig. 1:
\begin{eqnarray}
\rho_{xx}&\propto& C_{1}\cos w\tau +0.8C_{2}\cos 2w\tau
+0.2C_{3}\cos3w\tau+\nonumber\\
&& C_{1}[0.4\sin 2w\tau +0.2\sin 3w\tau +0.2\sin 4w \tau ]
\end{eqnarray}
For the middle panel:
\begin{eqnarray}
\rho_{xx}&\propto& C_{1}\cos w\tau +0.4C_{2}\cos 2w\tau
+0.1C_{3}\cos3w\tau+\nonumber\\
&&C_{1}[0.2\sin 2w\tau +0.1\sin 3w\tau ]
\end{eqnarray}
And finally for the bottom panel with the lowest $E_{0} /w$ ratio we
recover the harmonic response\cite{ina}:
\begin{equation}
\rho_{xx}\propto C_{1}\cos w\tau
\end{equation}

 These figures illustrate how the
$\rho_{xx}$ profile presents increasing anharmonicity as the MW
power is also increased. Thus, it can be observed clearly the
anharmonicity features: distorted profile in the $\rho_{xx}$
oscillations, new resonance peaks at the subharmonics of the
cyclotron frequencies and finally it is also remarkable that the
main oscillations are shifted to lower magnetic fields. All this
features corresponds unambiguously to a slightly anharmonic
behavior, i.e., the system amplitude is not very large yet. However
when a non-linear system is driven with very large amplitude, new
vibrational phenomena appear, like vibrations in which the motion
only repeat itself after two o more driver periods leading the
systems finally into a chaotic regime\cite{linsay}. This latter case
is not consider in this letter.
\begin{figure}
\centering\epsfxsize=3.5in \epsfysize=4.0in
\epsffile{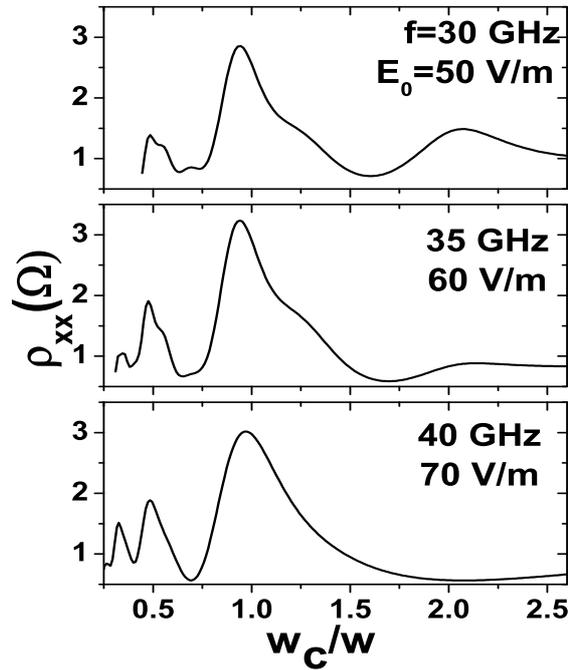}
\caption{Calculated $\rho_{xx}$ versus $w_{c}/w$ for three different
frequencies and MW intensities. We can observe how the anharmonicity
effects increase with decreasing frequencies. T=1K.}
\end{figure}

In Fig. 2 we present calculated $\rho_{xx}$ versus $w_{c}/w$ for
three different frequencies and MW intensities. We can observe, as
in the experiment\cite{doro},  how the anharmonicity effects
increase with decreasing frequency although the MW power also
decreases. This suggests that in the evolution to anharmonicity, $w$
plays a more important role than MW power. Thus, according to a
driven oscillating system, we propose that the ratio which govern
that evolution be proportional to the corresponding driven
amplitude:
\begin{equation}
\frac{eE_{o}}{m^{*}\sqrt{(w_{c}^{2}-w^{2})^{2}+\gamma^{4}}} \propto
\frac{E_{0}}{w^{2}}
\end{equation}

This work has been supported by the MCYT (Spain) under grant
MAT2005-06444, by the Ram\'on y Cajal program and  by the EU Human
Potential Programme: HPRN-CT-2000-00144.

%\clearpage

%Figure 1 caption: Calculated magnetoresistivity $\rho_{xx}$ as a
%function of $B$ for different MW intensities. MW frequency is $17
%GHz$. Vertical dashed lines correspond to the position of the
%cyclotron resonance ($w=w_{c}$) and its subharmnonics. It can be
%observed very clearly the increasing anharmonicity of $\rho_{xx}$ as
%the MW power is getting larger. T=1K.
%\newline

%Figure 2 caption: Same as in Fig. 1: calculated $\rho_{xx}$ versus
%$B$ for increasing MW power and $w=17 GHz$. In this case all results
%are displayed in the same panel.
%\newline

%Figure 2 caption: Calculated $\rho_{xx}$ versus $w_{c}/w$ for three
%different frequencies and MW intensities. We can observe how the
%anharmonicity effects increase with decreasing frequencies. T=1K.

\end{document}